\documentclass[useAMS,usenatbib]{mn2e}

\usepackage{natbib, aas_macros}
\usepackage[dvips]{graphicx}
\usepackage{wrapfig}
\graphicspath{{./figs/}}
\usepackage{color}
\usepackage{amsmath}
\usepackage{amssymb}

\newcommand{\Msun}{M_{\odot}}
\newcommand{\Mtot}{M_{\rm tot}}
\newcommand{\Zsun}{Z_{\odot}}
\newcommand{\fesc}{f_{\rm esc}}

\newcommand{\art}{\rm ART^{2}}
\newcommand{\A}{\rm \AA}
\newcommand{\La}{L_{\alpha}}

\newcommand{\lya}{\rm {Ly{\alpha}}}

\newcommand{\Msunyr}{\rm M_{\odot}~ yr^{-1}}
\newcommand{\fescion}{f_{\rm esc}^{\rm ion}}
\newcommand{\fsfr}{f_{\rm SFR}}
\newcommand{\tsfr}{\tau_{\rm SFR}}
\newcommand{\Rvir}{R_{\rm vir}}

\title[Effects of photon trapping on the $\lya$ properties of star-forming galaxies]
{Effects of photon trapping on the $\lya$ properties of star-forming galaxies}
\author[Yajima et al.]
{Hidenobu Yajima$^{1, 2, 3}$\thanks{E-mail: yajima@roe.ac.uk(HY);}
and Yuexing Li$^{2, 3}$
\\
$^{1}$ SUPA\thanks{Scottish Universities Physics Alliance}, 
Institute for Astronomy, University of Edinburgh, Royal Observatory, Edinburgh, EH9 3HJ, UK\\
$^{2}$Department of Astronomy and Astrophysics, Pennsylvania State University,
525 Davey Lab, University Park, PA 16802, USA\\
$^{3}$Institute for Gravitation and the Cosmos, The Pennsylvania State University, University Park, PA 16802, USA
}
 
\begin{document}

\date{Accepted ?; Received ??; in original form ???}

\pagerange{\pageref{firstpage}--\pageref{lastpage}} \pubyear{2008}

\maketitle

\label{firstpage}

%
%
\begin{abstract}

Recent observations show that a large number of Lyman-alpha emitters (LAEs) at high redshift $z \gtrsim 3$ have unusually high $\lya$ equivalent widths ($\rm EW \gtrsim 400 \; \A$). However, the origin of these high EWs is an open question. Here, we investigate the impacts of photon trapping on the $\lya$ EW and other properties by tracking the $\lya$ radiative transfer in spherical galactic clouds. We find that the delayed escape of the $\lya$ photons can change the $\lya$ properties significantly. During the transition phase from optically thick to optically thin where the $\lya$ photons can escape simultaneously, the EW can be boosted to $\sim 1000 \; \A$, the $\lya$ luminosity can be increased by a factor of a few, and the line profile can be significantly broadened. The boost factor appears to depend on the galaxy properties such as mass and star formation rate and timescale, therefore future investigation combing 3D $\lya$ RT calculations with cosmological simulations of galaxy formation and evolution is needed to fully understand the $\lya$ properties of early star-forming galaxies. 

\end{abstract}

%
%

%
%
\section{Introduction}

The hydrogen $\lya$ line has played an important role in detecting distant galaxies. To date, a large population of $\lya$ emitters (LAEs) have been observed at high redshifts $z \gtrsim 3$ \citep[e.g.,][]{Hu96, Cowie98, Iye06, Gawiser06, Gronwall07, Lai08, Ouchi08, Lehnert10, Kashikawa2012, Shibuya12}. One of the important properties of $\lya$ is the equivalent width (EW), which measures the ratio of $\lya$ flux to that of the UV continuum, and therefore provides useful information of the emission mechanisms and photon escape processes in galaxies. 

Unlike UV continuum, $\lya$ photons typically experience many scattering process in galaxies due to its large scattering cross section. These scatterings may prolong the path length and increase the probability of dust absorption, leading to reduction in the emergent luminosity. Hence, the EW can change significantly due to the difference in the radiative transfer (RT) process between $\lya$ and UV continuum. If the escape fraction of $\lya$ is the same as that of the UV continuum, the upper limit of the EW is determined by the intrinsic spectral energy distribution (SED) of the stars. It has been suggested that in the case of a Salpeter initial mass function (IMF), the upper limit is $\sim 400~\A$ \citep[e.g.,][]{Schaerer03}.

However, a large number of LAEs at $z \gtrsim 3$ have large EWs higher than $400~\A$ \citep[e.g.,][]{Gronwall07, Ouchi08}. More recently, \cite{Kashikawa2012} detected a LAE with extremely large EW of $\sim 900~\A$ at $z=6.5$. The origin of such high EWs is largely unknown. 

It has been proposed that the large EWs may be produced by a top-heavy IMF of PopIII stars \citep[e.g.,][]{Gronwall07, Kashikawa2012}. However, so far there is no evidence of a top-heavy IMF in high-redshift LAEs \citep[e.g.,][]{Nagao08}, or a PopIII-dominated galaxy. Another model used a Salpeter IMF but invoked clumpy interstellar medium \citep{Neufeld91, Hansen06}. In this model, UV continuum is assumed to be absorbed by dust passing through the dense clouds, but $\lya$ photons do not experience the dust absorption because they cannot enter the clouds due to surface scattering. However, this is an idealized, optically thin case for the $\lya$ photons in emitting regions between the dense clouds. 
In fact, most of the stars form in high density clouds, they ionize the gas and create $\rm H_{II}$ regions which emit $\lya$ photons.  These photons then travel in the dense clouds and may experience strong dust extinction. Recently, \cite{Yajima12d} studied the escape of $\lya$ and continuum photons from star-forming galaxies by combining a cosmological simulation with three-dimensional radiative transfer calculations, and found no correlation between the EW and clumpiness of the ISM. 
In addition, Laursen et al. (2012) studied the EWs of modeled galaxies with clumpy ISM, 
and showed that it was difficult to boost the EW with the typical physical state of LAEs. 
On the other hand,
\cite{Yajima12c} found that collisional excitation from accretion of cold hydrogen gas by galaxies at high redshift ($z \gtrsim 6$) significantly enhances the $\lya$ emission and boost the EW. 

It is not clear, however, from our previous studies whether or not cold accretion is the only dominant mechanism to produce a large EW. In many RT calculations,  all $\lya$ photons were assumed to escape instantaneously with the UV continuum. However, since the ionization of hydrogen evolves with  time, this time-dependent propagation of the ionization front may cause a delay of the escape of the $\lya$ photons \citep{Pierleoni09}, which in turn may affect the $\lya$ properties. However, the time evolution of ionization has not been considered in most $\lya$ RT calculations before.

In this work, we consider the pure effects of photon trapping on the $\lya$ properties. We perform full $\lya$ RT, which includes time-evolving ionization structure, in spherical galactic clouds, and track the resulting $\lya$ EW, luminosity, and line profiles. The paper is organized as follows:  in \S~2 we describe our galaxy models and RT methods, in \S~3 we present the results of the $\lya$ properties, we discuss   our model assumptions and  parameters in \S~4, and summarize in \S~5.

%
%
\section{Methodology}

\subsection{Galaxy Models}

Our galaxy model consists of one-dimensional spherical shells of hydrogen gas in a spherical Navarro-Frenk-White dark matter halo \citep{NFW}, similar to that of \citet{Dijkstra09}. The dark matter halo has a  concentration parameter $c = 3.8$ \citep{Gao08} at the virial temperature of the halo as derived by \citet{Makino98} at $z_{\rm vir} = 3$.  The gas is assumed to be in hydrostatic equilibrium with the virial temperature and then gets cold isothermal state 
with a temperature of $T = 10^{4}$, as suggested by
\citet{Dijkstra09}, who showed that gas in galaxies can quickly reach the cold state of $\sim 10^{4}$ K as a result of balancing between gravitational-shock heating and radiative cooling.
A total of 100 shells are constructed for each galaxy.

In addition, we assume star formation takes place at the center of the galaxy, with the rate SFR proportional to the halo mass \citep[e.g.,][]{Trac07},
and decays exponentially,

\begin{equation}
{\rm SFR}(t) = f_{\rm SFR}\; \left( \frac{M_{h}}{10^{10}\; \Msun} \right) \;{\rm exp} \left(- \frac{t}{\tsfr} \right)
\end{equation}
where $f_{\rm SFR}$ and $\tsfr$ are amplitude factor of peak SFR and star formation timescale respectively. 
In this work, the $f_{\rm SFR}$ and $\tsfr$ are free parameters with a range $f_{\rm SFR} = 0.05 - 5.0$ and  $\tsfr = 5 \times 10^{6} - 10^{8}$ yr.
As fiducial models, we choose $\Mtot = 10^{10}$ and $10^{11}~\Msun$ with $\tsfr = 5 \times 10^{7}$ and 
$f_{\rm SFR} = 0.5$ (the peak SFR ranges $0.5 - 5~\Msunyr$), as suggested by observations of LAEs from both clustering analysis \citep[e.g., ][]{Gawiser07, Ouchi08, Ouchi10, Guaita10} and SED fitting \citep{Gawiser06, Gawiser07, Finkelstein09, Nakajima12}.

Along with the star formation history, newly formed stars are added to the galaxy center. We estimate the number of ionizing photons from stacked SEDs.
The SED is calculated by summing that star clusters are created in each time step. We use PEGASE ver.2.0 \citep{Fioc97} to calculate the SEDs, and 
 assume a Salpeter initial mass function with a metallicity of $Z = 0.05 \Zsun$. 
In each time step, we estimate the ionization structure from the balance between recombination and photo-ionization,
\begin{equation}
\dot{N}_{\rm ion} = \int_{0}^{R_{\rm i}} 4 \pi r^{2} \alpha_{\rm B} n_{\rm HII}(r) n_{\rm e}(r) dr
\end{equation}
where $\dot{N}_{\rm ion}$, $R_{\rm i}$, $\alpha_{\rm B}$ are the emissivity of ionizing photons, the radius of ionized region and the recombination rate to all excitation state, respectively. The gas at $r \le R_{\rm i}$ is completely ionized region, while at $r > R_{\rm i}$ it is in neutral state.

\subsection{Radiative Transfer}
\label{sec:radiation}

The RT calculations are performed using the 3D Monte Carlo RT code, All-wavelength Radiative Transfer with Adaptive Refinement Tree ($\art$) \citep{Li08, Yajima12b}. The multi-wavelength continuum emission of $\art$ is described in \cite{Li08}, while the implementation of $\lya$ line transfer is detailed in \cite{Yajima12b}. Here we focus on the RT of $\lya$ photons in the ionization structure of the galactic shells and briefly outline the process. 
 
In the $\lya$ RT, it is important to calculate the frequency change due to the scattering process. In the scattering process, the final frequency in the laboratory frame is then:
\begin{equation}
\frac{\nu^{\rm out} - \nu_{0}}{\Delta \nu_{\rm D}} = \frac{\nu^{\rm in} - \nu_{0}}{\Delta \nu_{\rm D}}
- \frac{V_{\rm a} \cdot k_{\rm in}}{V_{\rm th}} + \frac{V_{\rm a} \cdot k_{\rm out}}{V_{\rm th}}
\end{equation}
where, $\nu^{\rm in}$ and $\nu^{\rm out}$ are incoming and outgoing frequency in the rest frame of scattering medium, respectively,
the line center frequency $\nu_{0} = 2.466 \times 10^{15}$ Hz, 
$\Delta \nu_{\rm D} = (V_{\rm th} / c) \nu_{0}$ corresponds to the  Doppler frequency width and
$V_{\rm th}$ is the velocity dispersion of the Maxwellian distribution describing the thermal motions,
i.e., $V_{\rm th} = (2 k_{\rm B} T / m_{\rm H})^{1/2}$,
$V_{\rm a}$ is the atom velocity,
$k_{\rm in}$ and $k_{\rm out}$ are incoming and outgoing propagation direction respectively.

Unlike continuum photons, the $\lya$ photons can experience many scattering processes, which lead to long path length and long scattering timescale before the photons escape. As a result, the ionization structure may change while the photons are traveling. Here, we consider the change of ionization structure with  time in the $\lya$ RT process. The optical depth is estimated to be $d\tau = \sigma_{\nu} n_{\rm HI} (i, j) ds$, where $i$ and $j$ are indexes of the shell position and the time step, respectively. In each scattering, if the traveling of the $\lya$ moves to the next time step, the ionization structure of that time step is used to estimate the optical depth.

The intrinsic $\lya$ luminosity is estimated as:
\begin{equation}
L_{\alpha} = 
0.68 h \nu_{\alpha} \dot{N}_{\rm ion} (1 - f_{\rm esc}^{\rm ion})
\end{equation}
 where the constant 0.68 is the rate of transition
  from 2p to 1s state via the recombination process under the approximation of case B (Osterbrock 2006),
and $\nu_{\alpha}$ and $f_{\rm esc}^{\rm ion}$ are the $\lya$ frequency and the escape fraction of ionizing photons, respectively.
The starting point of the photon package is randomly chosen with the weight by the $\lya$ emissivity of each spherical shell,
$\epsilon_{\alpha}^{i} \propto \pi(r^{3}_{i} - r^{3}_{i-1})n_{\rm HII}^{i}n_{\rm e}^{i}$, where $i$ is the shell index.

A total of 128 photon packets are used at each time step. Each packet has
the same energy as the $\La$ at the time step divided by the
total packet number.
The initial frequency is randomly chosen from a Gaussian distribution.


%
%

\section{Results}

\subsection{Ionization History}

\begin{figure}
\begin{center}
\includegraphics[scale=0.4]{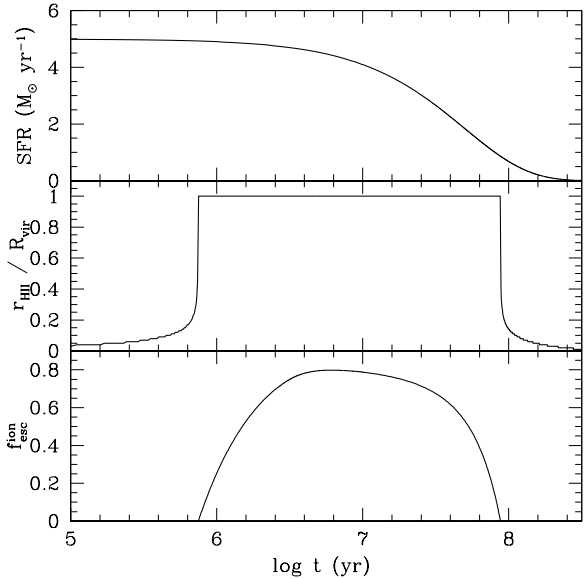}
\caption{
Time evolution of the SFR ({\it upper panel}), size of the ionization front
normalized by the virial radius ({\it middle panel}), and the escape fraction
of ionizing photons ({\it lower panel}), for a galaxy with mass
$\Mtot = 10^{11} ~\Msun$, and fiducial parameters with star formation timescale $\tsfr = 5\times10^{7}$ yr, and star formation amplitude factor $\fsfr = 0.5$. 
}
\label{fig:sfr}
\end{center}
\end{figure}

Figure~\ref{fig:sfr} shows the evolution of the star formation rate, the size of the ionized region, and the escape fraction of ionizing photons $\fescion$ with time for a galaxy with mass $\Mtot = 10^{11} ~\Msun$, a star formation timescale $\tsfr = 5\times10^{7}$ yr, and a star formation amplitude factor $\fsfr = 0.5$. The SFR has an initial value of $5 \; \Msunyr$, but exponentially decreases with time.  Since ionizing photons mainly come from massive, young stars, they are strongly correlated with SFR. As a result, in the early phase ($t \lesssim 9 \times 10^{7}$ yr), the ionizing front reaches the  virial radius ($\Rvir$).
Then, the ionization front steeply decreases with SFR, and becomes $\sim 0.1 R_{\rm vir}$ at $\sim 10^{8}$ yr. 

In the current model of ionization, the $\fesc$ of ionizing photons becomes $> 0$ when all gas in the galaxy is ionized. Therefore, the $\fesc$ is nonzero at $t \lesssim 9\times 10^{7}$ yr. It starts from $\sim 0.8$, then decreases with SFR. When all gas is ionized, the $\fesc$ is estimated by $1 - \dot{N}_{\rm rec, tot}/ \dot{N}_{\rm ion}$, where $\dot{N}_{\rm rec, tot}$ is the number of total recombination hydrogen, 
i.e., $\dot{N}_{\rm rec, tot} = \int_{0}^{\Rvir} 4 \pi r^{2} \alpha_{\rm B} n_{\rm HII}(r) n_{\rm e}(r) dr$. 
In this case, $\dot{N}_{\rm rec, tot}$ is time-independent, and the $\fesc$ is correlated with SFR.

\subsection{$\lya$ Equivalent Width}

\begin{figure}
\begin{center}
\includegraphics[scale=0.4]{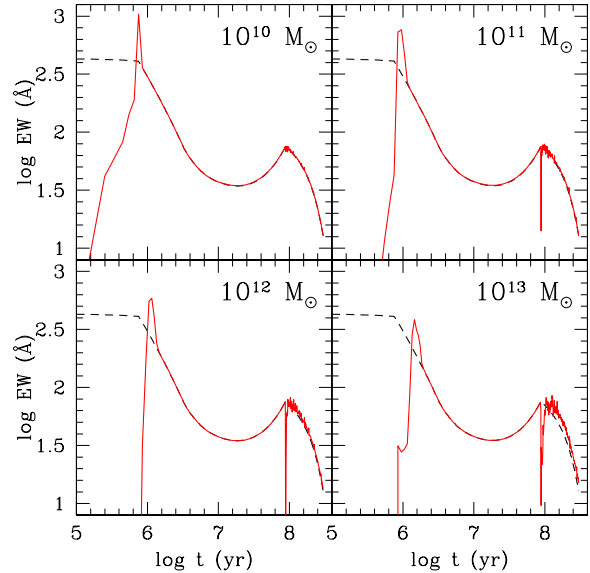}
\caption{Time evolution of the $\lya$ EW for 4 different galaxies with mass of $10^{10}\, \Msun$ (upper left panel), $10^{11}\, \Msun$ (upper right panel), $10^{12}\, \Msun$ (lower left panel), $10^{13}\, \Msun$ (lower right panel), respectively. Fiducial parameters $f_{\rm SFR} = 0.5$ and $\tau_{\rm SFR} = 5 \times 10^{7}$~yr are used for all galaxies. The red solid line is the resulting EW from our model which takes into account the trapping of $\lya$ photons due to the time evolution of ionization structure in the galaxy, while the black dashed line is the predicted EW assuming that all $\lya$ photons escape instantaneously. A Salpeter IMF is assumed in these calculations.
}
\label{fig:mass_comp}
\end{center}
\end{figure}

We estimate the EW in rest frame from $EW = \La / L_{\lambda}^{\rm UV}$,  where $L_{\lambda}^{\rm UV}$ is luminosity density of UV continuum, here we use the luminosity density at $1400~\A$. Figure~\ref{fig:mass_comp} shows the resulting time evolution of EWs with the fiducial values $\fsfr = 0.5$ and $\tsfr = 5 \times 10^{7}$ yr, for 4 different galaxies in the mass range of $10^{10-13}\, \Msun$. 

At first, the ionized region is confined in the galaxy, then it grows with increasing stellar mass. In this phase, the $\lya$ photons are also confined in the galaxy,
hence the EW is very small. However, when the ionization front reaches the virial radius, the trapped $\lya$ photons can escape simultaneously. This sudden release of $\lya$ photons causes very high EWs, which peaks $\sim 400 - 1000~\A$, much higher than the upper limit predicted by 
the model in which all $\lya$ photons escape instantaneously.

Moreover, the peak of EW decreases with increasing halo mass. 
In low-mass galaxies, the potential well is shallow, 
and the virial radius $\Rvir$ is small,  
therefore it is easy for the ionizing photons to propagate and reach $\Rvir$ quickly. 
 Since the ionization timescale of small galaxies 
  is shorter than that of massive ones, by the end of ionization, a larger
  ratio of massive ($> 10~\rm \Msun$) to intermediate-mass stars may result
  due to the short lifetimes of massive stars. This would lead to larger EWs
  because more massive stars produce more $\lya$ emission via ionizing photons, while
  intermediate- or low-mass stars (which have long lifetimes) mainly
  contribute UV continuum. Therefore, small galaxies may produce larger EWs
  than massive ones.
In addition, in the larger radius, the traveling time from last scattering to escape can change depending on the position and angle, resulting in the extended and smaller peak of EW.

Once the galaxy is ionized, the ionizing photons can escape, and the total recombination rate is fixed. 
From Equations (2) and (4), the intrinsic $\lya$ luminosity then becomes constant, $\La = 0.68 h\nu_{\alpha}\int_{0}^{R_{\rm vir}} 4 \pi r^{2} \alpha_{\rm B} n_{\rm H}^{2}(r) dr$. On the other hand, the UV continuum can continue to increase with the formation of new stars. Thus, upon completion of ionization, the intrinsic $\lya$ luminosity decreases relatively to UV continuum, resulting in a decrease of $\lya$ EW. 

After that, however, the ionized region is confined within $\Rvir$ again due to decreased SFR. Most of $\lya$ photons are trapped within the galaxy by the neutral hydrogen shell. Some of the trapped photons can escape through scattering and random processes, leading to a mild increase of the EW. However, the UV continuum slowly catches up again with new star formation, bringing down the EW.  For galaxies above $10^{11}~\rm \Msun$, there are sharp dips at $t \sim 10^{8}~\rm yr$.
This is the phase transition from complete to partial ionization, i.e., the
gas shell of neutral hydrogen forms again. At the transition, $\lya$ photons start to be trapped by the shells, 
hence the EW decreases steeply. For the galaxy of $10^{10}~\rm \Msun$,
however, due to the smaller size, the photon traveling time is much smaller than the timestep of
simulations, resulting in negligible decrease. 

\begin{figure*}
\begin{center}
\includegraphics[scale=0.6]{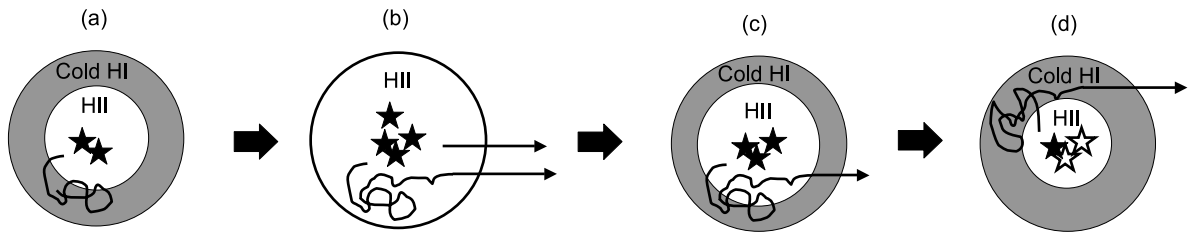}
\caption{
Schematic view of the interplay between ionization and $\lya$ escaping as a
function of time in a galaxy in which the hydrogen gas is
initially settled in cold state of $10^{4}$ K.
(a):  the ionized region is confined in the galaxy, the $\lya$ photons are trapped.
(b):  the galaxy is completely ionized, the accumulated $\lya$ photons escape instantaneously.
(c):  the ionized region is confined again, and the $\lya$ photons are trapped again. 
(d):  the stars become older and dimmer, some $\lya$ photons escape through scattering. 
}
\label{fig:idea}
\end{center}
\end{figure*}

The interplay between $\lya$ and UV continuum photons in the RT processes results in a unique saddle-like shape of the EW as a function of time. To better understand the oscillatory behavior of the EW, we illustrate in Figure~\ref{fig:idea}  a schematic cartoon of the time evolution of the ionization structure and the escape of $\lya$ photons. There are four major phases in the evolution which affect the $\lya$ EW significantly: 

\begin{description}
\item[(a):]  Early phase when the ionized region is confined in the galaxy due to low SFR and stellar mass. Most of the $\lya$ photons are trapped in the galaxy. The resulting $\lya$ EW is very small.
\item[(b):]  As more stars form, the flood of ionizing photons from young, massive stars quickly ionizes the galaxy. The previously trapped $\lya$ photons as well as the new ones created via recombination of the ionizing photons in the HII regions escape instantaneously, causing a sudden and strong boost of the EW. 
\item[(c):]  With the decline of SFR due to ionization, the number of ionizing photons is reduced, and the ionized region is confined in the galaxy again. The cooled and neutral hydrogen can trap the $\lya$ photons, leading to a decrease of the EW.
\item[(d):] As the stars become older and dimmer, the UV continuum flux decreases. Meanwhile, some of the trapped $\lya$ photons escape after some scatterings. Hence, the $\lya$ luminosity increases relatively to the UV continuum,  bringing a rise of  the EW again.
\end{description}

\subsection{$\lya$ Luminosity}

\begin{figure}
\begin{center}
\includegraphics[scale=0.4]{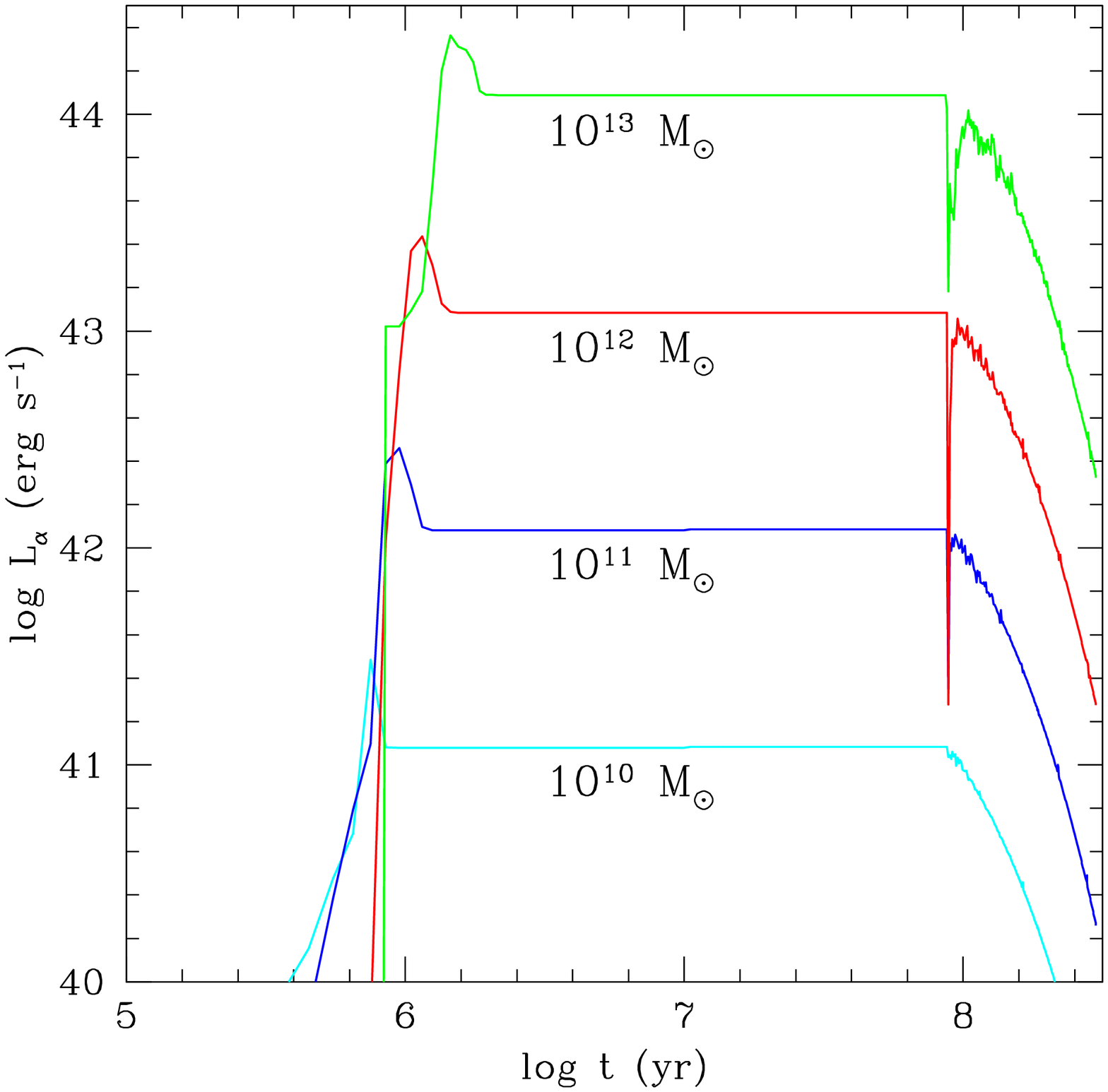}
\caption{Evolution of emergent $\lya$ luminosity from the four modeled galaxies as indicated by the colors 
(cyan : $\Mtot = 10^{10}~\Msun$, blue : $\Mtot = 10^{11}~\Msun$, 
 red : $\Mtot = 10^{12}~\Msun$ and green : $\Mtot = 10^{13}~\Msun$).
}
\label{fig:Lcomp}
\end{center}
\end{figure}

Owing to the time-dependence of the ionization structure, the emergent $\lya$ luminosity ($\La$) from the galaxy evolves with time as well. Figure~\ref{fig:Lcomp} shows the corresponding $\La$ from the modeled galaxies as a function of time. 
 
During Phase (a), the $\lya$ photons are trapped in the galaxy, the $\La$ approaches to zero in the modeled galaxies. As the galaxy gradually becomes optically thin due to ionization, $\lya$ photons can escape, steadily increasing the luminosity. Upon Phase (b) when the galaxy is completely ionized, all the $\lya$ photons, including both the previously trapped and the newly created ones, escape instantaneously, resulting in a drastic, strong peak of the $\La$. After that, the ionized region is confined in the galaxy, the $\lya$ photons are trapped again as in Phase (c) and (d). But since the $\lya$ photons escape mainly through scattering, the fraction remains nearly constant during these phases, leading to constant luminosity.  At the end of the star formation timescale, no more new stars form, the number of ionizing and $\lya$ photons declines sharply, so the galaxy becomes optically thick, and most of the $\lya$ photons become trapped, resulting in a rapid drop of the luminosity. 

The evolution of the $\lya$ luminosity does not show the characteristic "saddle" shape as the EW. This is because the $\lya$ luminosity depends only on the ionization structure and the escaping process, it does not depend strongly on the UV continuum as does the EW. 

In contrast to the EW, which in general decreases with galaxy mass, the $\lya$ luminosity increases with galaxy mass. This is because both the SFR and the total recombination number are proportional to the galaxy mass, and as a result, the $\La$ increases linearly with the galaxy mass.

\subsection{$\lya$ Line Profiles}

\begin{figure}
\begin{center}
\includegraphics[scale=0.4]{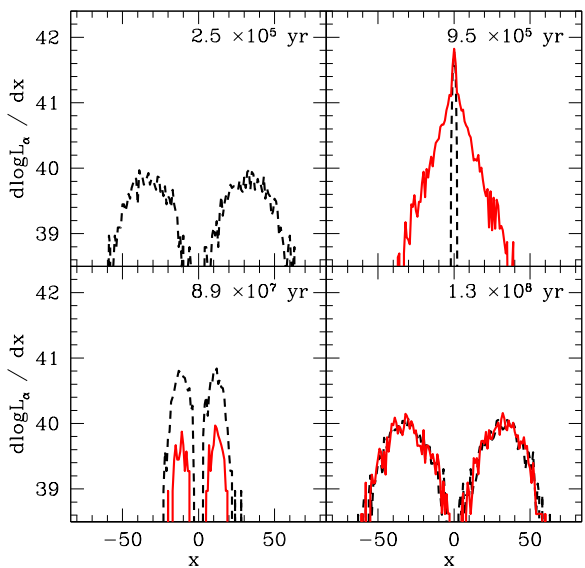}
\caption{
The $\lya$ line profile of the $\Mtot = 10^{11} ~\Msun$ galaxy at four different times: $t = 2.5 \times 10^{5}, 9.5 \times 10^{5}, 8.9 \times 10^{7}$ and $1.3 \times 10^{8}$ yr, respectively. Fiducial parameters star formation timescale $\tsfr = 5\times10^{7}$ yr, and star formation amplitude factor $\fsfr = 0.5$ are used. 
The red solid line is the line profile from our model which takes into account the trapping of $\lya$ photons due to the time evolution of ionization structure in the galaxy, while the black dashed line is the predicted profile assuming that all $\lya$ photons escape instantaneously. The $x$ is frequency shift in Doppler units, $x = (\nu - \nu_{0}) / \Delta \nu_{\rm D}$.
}
\label{fig:profile}
\end{center}
\end{figure}

Figure~\ref{fig:profile} shows the emergent line profiles from the model which takes into account the delayed escape of $\lya$ photons due to the time-dependence of the ionization structure in the galaxy, in comparison with those without photon trapping. 

At $t = 2.5 \times 10^{5}$ yr, which corresponds to Phase (a) in Figure~\ref{fig:idea}, the delayed-escaping model predicts that the $\lya$ photons are trapped in the galaxy, there is no emergent line emission. In contrast, the instantaneous-escaping model predicts a double-peak profile caused by scattering of the neutral shells.   At Phase (b) when the galaxy is completely ionized ($t = 9.5 \times 10^{5}$ yr), the galaxy becomes optically thin. The line profile without the time dependence shows a narrow profile peaking around $x=0$, which is the intrinsic profile, while the line from the trapping model shows not only the central peak, but also a significantly extended distribution in both the blue and red wings. The broadening is caused by the scatterings of the trapped photons before they escape. Then, in Phase (c) ($t = 8.9 \times 10^{7}$ yr), the galaxy becomes optically thick again, the line profile thus shows double peak owing to scattering, and the luminosity from the trapping model is lower than that from the no-trapping model as expected. In Phase (d) ($t \sim 1.3 \times 10^{8}$ yr), with the decline of SFR, the size of the ionized region becomes small. As a result, the optical depth from the center to the $\Rvir$ increases, the distance between the peaks in the profile thus increases, resulting in two broad double peaks.

%
%

\section{DISCUSSIONS}

\begin{figure}
\begin{center}
\includegraphics[scale=0.4]{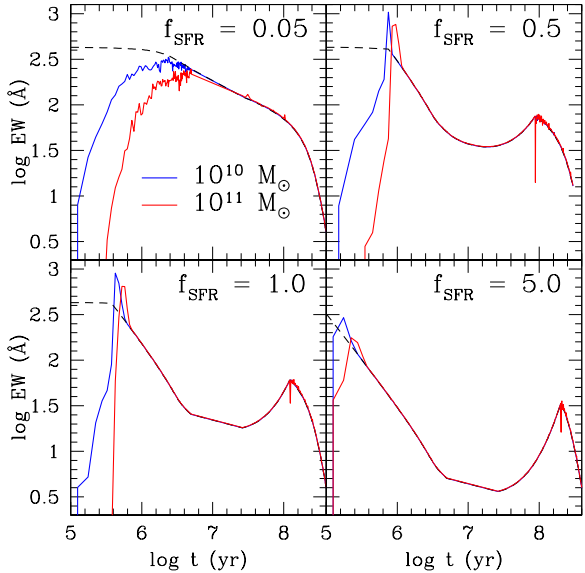}
\caption{
The evolution of EWs from two galaxies with mass of $10^{10}~\Msun$ and
$10^{11}~\Msun$, respectively, at $\tau_{\rm SFR} = 5 \times 10^{7}$ yr but
different amplitude factor of SFR. 
Black dash lines are the EW in the assumption that all $\lya$ photons instantaneously escape.
The solid lines are the EW considering the time evolution of ionization structure while $\lya$ photons travel in the galaxy.
}
\label{fig:sfr_comp}
\end{center}
\end{figure}

\begin{figure}
\begin{center}
\includegraphics[scale=0.4]{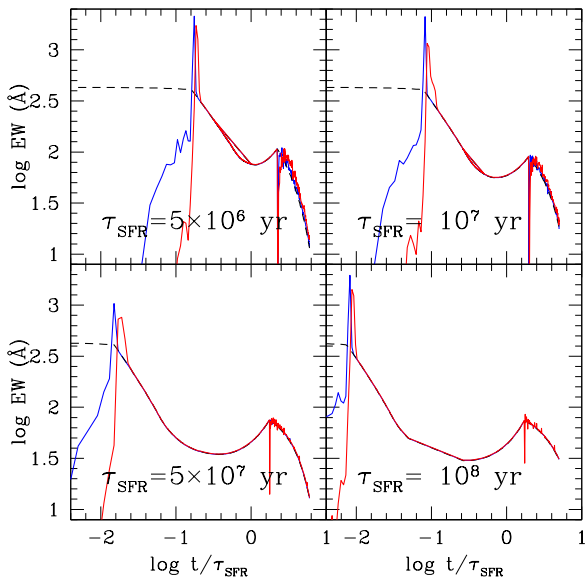}
\caption{
The evolution of EWs from two galaxies with mass of $10^{10}~\Msun$ and
$10^{11}~\Msun$, respectively, at $f_{\rm SFR} = 0.5$ but different star
formation timescales. 
Black dash lines are the EW in the assumption that all $\lya$ photons instantaneously escape.
Solid lines are the EW considering the time evolution of ionization structure while $\lya$ photons travel in the galaxy.
Same as in Figure~\ref{fig:sfr_comp}, the blue curve represents the $10^{10}~\Msun$ galaxy, while the
  red curve represents the $10^{11}~\Msun$ galaxy.
 }
\label{fig:time_comp}
\end{center}
\end{figure}

In our model, there are two free parameters, the amplitude factor of peak SFR, $f_{\rm SFR}$, and star formation timescale,  $\tsfr$. In the previous sections, we have presented results using the fiducial numbers $\fsfr = 0.5$ and $\tsfr = 5\times10^{7}$ yr. Here we explore the dependence of our results on these two parameters. 

Figure~\ref{fig:sfr_comp} shows the EWs of two galaxies of $\Mtot =
10^{10}~\Msun$ and $10^{11}~\Msun$, respectively, with the same star formation timescale
$\tsfr = 5\times 10^{7}$ yr but different star formation factor $\fsfr$.
For $\fsfr = 0.05$, the ionized region is always confined in the galaxy owing to low SFR. Hence, there is no EW peak even at the Phase (b) in figure~\ref{fig:idea}. The EW simply decreases with SFR as the mean stellar age becomes old. For $\fsfr \sim 0.5 - 1.0$, there is a EW peak at Phase (b). With higher $\fsfr$, however, both the SFR and the mass of young stars are higher, leading to an earlier completion of ionization and higher UV continuum flux, which results in smaller EW.

Figure~\ref{fig:time_comp} shows the EWs of two galaxies of $\Mtot =
10^{10}~\Msun$ and $10^{11}~\Msun$, respectively, with the same star formation factor $\fsfr = 0.5$ but different star
formation timescale $\tsfr$. 
All plots show significantly boosted EW peak at Phase (b), and similar ``saddle'' shape of the evolution of EW. The difference among these models lies in the duration of Phase (c). With shorter $\tsfr$, the the occurrence of Phase (b) is more significantly delayed, resulting in larger difference in the EW evolution between the trapping and non-trapping models. 

These results suggest that the EW depend sensitively on the galaxy properties such as star formation rate and timescale. In order to fully address the origin and nature of the EW distribution in LAEs, it is desired to combine 3D $\lya$ RT calculations with full cosmological simulations which include more realistic treatments of star formation and feedback processes, which we plan to do in future work. 
In addition, 3D hydrodynamics and RT calculations would produce more realistic
temperature structures. 
Here, we assume all gas to be $10^{4}$ K due to efficient radiative cooling. 
However, at outer low-density region, some fraction of gas can maintain high temperature near the virial temperature after gravitational shock. 
Such a detailed history of thermal evolution may change the ionization
structure via collisional ionization process, which may lead to different evolution of EW. 

In our galaxy model, LAEs are constructed as spherical clouds
  with embedded stars. If the morphological shape is disky or irregular,
  supernovae or radiative feedback can produce asymmetric gas ionization and
  outflow, which would facilitate escaping of ionizing photons. It was
  suggested that dense gas shells may still form and efficiently trap photons
  in the presence of galactic outflows in dwarf galaxies
  \citep{Fujita03}. However, we note that the relation between $\lya$ EW and
  outflow gas velocity is a complicated issue as the EW depends sensitively on
  a number of factors such as ionization structure, dust content and
  metallicity \citep{Yajima12d}. LAEs typically have high EWs and low gas
  outflow velocities, while Lyman Break Galaxies (LBGs), which are typically
  more massive and metal rich than LAEs, appear to have low EWs but high
  outflow velocities, as suggested by recent observations \citep[]{McLinden11, Finkelstein11, Hashimoto13}. This may explain the ``anti-correlation" between $\lya$ EW and velocity offset reported by \citet{Hashimoto13} (as well as by Erb et al. 2013, in prep) in galaxy samples that contain both LAEs and LBGs. We plan to investigate the relation between $\lya$ EW and outflow gas velocity using a larger galaxy population that spans a wider range of $\lya$ and properties from cosmological simulations in the future.

In this work, we did not consider dust extinction and detection threshold of flux in observation. 
The dust extinction can reduce the EW due to smaller escape fraction of $\lya$ photons. 
However, recent observation indicated that the dust extinction of
$\lya$ photons might be similar as that of UV continuum. 
If this is the case, the resulting EW may not be reduced significantly by the dust. 
On the other hand, imposing a detection threshold may reduce the EW here
because we assume the flux is contributed by all $\lya$ photons. In regions
with extended $\lya$ emission due to the scattering processes, the surface
brightness may be below the survey threshold and not be detected. We will 
include more realistic treatments of these issues in future work using cosmological simulations.

%
%

\section{SUMMARY}

We have performed a set of $\lya$ RT calculations on idealized spherical galaxies to study the effects of photon trapping on the $\lya$ properties in star-forming galaxies, by taking into account the time dependence of the ionization process. We have identified four major phases in the evolution of the galaxy in our model:

\begin{itemize}
\item Phase (a):  The ionized region is confined in the galaxy due to low SFR and stellar mass. Most of the $\lya$ photons are trapped in the galaxy, no emergent $\lya$ emission is observed.
\item Phase (b):  The galaxy is completely ionized. The previously trapped $\lya$ photons as well as the new ones created via recombination escape instantaneously, resulting in a dramatic and strong boost of the EW and luminosity, and a broadened line profile.  
\item Phase (c):  The ionized region is confined in the galaxy again due to decrease of SFR and ionizing photons. The $\lya$ photons are trapped again, leading to a decrease of the EW and luminosity.
\item Phase (d): The ionized region becomes smaller, as the stars become older and dimmer. The escape of the trapped $\lya$ photons is dominated by the stochastic scattering process. There is a small enhancement of the EW due to the decrease of the UV continuum flux, but the $\lya$ luminosity remains constant until the galaxy becomes strongly optically thick and trap most of the photons.
\end{itemize}

Our model suggests that the interplay between time-dependent ionization and $\lya$ scattering impacts significantly the $\lya$ EW, the emergent luminosity, and the line profile. During the transition phase from optically thick to optically thin where the $\lya$ photons can escape simultaneously, the EW can be boosted to $\sim 1000 \; \A$, the $\lya$ luminosity can be increased by a factor of a few, and the line profile can be significantly broadened. These results may be able to explain the unusually large EW in some observed LAEs. However, the boost factor depends on the galaxy properties such as mass and star formation rate and timescale, future investigation combing 3D $\lya$ RT calculations with cosmological simulations of galaxy formation and evolution is needed to fully address this issue.

%
%
\section*{Acknowledgments}
We thank  Masakazu A. R. Kobayashi, Tom Abel and Masayuki Umemura for
stimulating discussions and helpful comments, as well as the referee for a
constructive report which has helped improve the paper. 
Support from NSF grants AST-0965694 and AST-1009867 is gratefully acknowledged. We acknowledge the Research Computing and Cyberinfrastructure unit of Information Technology Services at The Pennsylvania State University for providing computational resources and services that have contributed to the research results reported in this paper (URL: http://rcc.its.psu.edu). The Institute for Gravitation and the Cosmos is supported by the Eberly College of Science and the Office of the Senior Vice President for Research at the Pennsylvania State University.

%
%




\label{lastpage}

\end{document}